\documentclass[12pt]{article}

\def\i{\mbox{i}}
\def\d{\mbox{d}}

\title{Integral in the sense of principal value \\
as a distribution over parameters of integration}
\author{M. L. Nekrasov \\
{\small\it Institute for High Energy Physics, 142284 Protvino,
Russia}}

\date{}

\begin{document}

\maketitle

\begin{abstract}
An integral in the sense of principal value of a singular function
or of product of singular functions can appear itself as a
singular function in some range of values of integration
parameters. In this case, if necessary subsequently to integrate
with respect to parameters, the problem arises about
interpretation of the initial integral as a distribution over the
integration parameters. A solution to this problem is offered,
which is initiated by actual applications in quantum field theory.
\end{abstract}

\section{Introduction}

In many applications of quantum field theory one has to deal with
integrals determined in improper sense. A classical example is the
integrals of Feynman diagrams where the propagators of particles
are determined (in Minkowski space) with a specific rule of the
bypass of mass-shell singularities. With the loop corrections are
switched on also the ultraviolet divergences appear in such
integrals. The consistent and mathematically verified solution to
the problem of ``elimination'' of the ultraviolet divergences is
based on the conception of the extending of linear continuous
functionals---what the coefficient functions of $S$-matrix are
\cite {BS}---from a class of rapidly decreasing functions onto a
class of arbitrary regular functions. As a matter of fact the
solution is based on applying the distribution-theory methods
\cite{Sobolev,S,GS,BLOT}. For the first time the idea of such an
approach for solving the problem of ultraviolet divergences was
put forward by N.N.Bogolyubov \cite{B1952} and afterwards was
realized by him in collaboration with O.S.Parasiuk \cite{BP}, and
further by many other authors (see bibliography and the detailed
account in monography \cite{BS}).

A characteristic feature of the solution to the problem of
ultraviolet divergences is the appearance of free finite
parameters in the theory (in the renormalizable theories they are
absorbed in favor of the renormalized constants). From the point
of view of theory of distributions the mentioned property is
entirely natural as it is connected with the operation of the
extending of linear continuous functionals. Nevertheless, in many
cases the solving by the same methods of other problems can be
realized with the subsequent elimination of the ambiguities.
Actually this occurs if the solution implies imposing of
additional condition(s). Well-known example is the definition of a
propagator in Minkowski space where the bypassing rule is fixed by
the causality condition. In specific field-theory applications the
elimination of ambiguities can be carried out by the imposing of a
self-consistency condition (see nontrivial example in~\cite{NR}).

Among the problems whose solution is based on the extending of
linear continuous functionals but is realized without the
emergence of ambiguities, a particular place is occupied by the
problem of asymptotic expansion in a parameter of an integral
defining some quantity (for example, the amplitude or probability
of a physical process). Really, if the expansion is carried out
before the calculation of the integral, then the expansion can led
to the appearance of singular functions in the integrand, which
are not integrable in the conventional sense. In this case for
giving a sense to divergent integrals one can take advantage of
the theory of the extending of linear continuous functionals (the
theory of distributions). However, in contrast to the case of
renormalizations, the ambiguities arising therewith must be
completely eliminated---as the initial integral before performing
the asymptotic expansion was well-determined (by our assumption)
and its expansion contained no ambiguities.\footnote{The problem
of asymptotic expansion under the symbol of integral arises when
the integral cannot be calculated for technical reasons, but an
expansion of the integral in a parameter is required.} A general
recipe of the elimination of the ambiguities is explained in
\cite{T}. Nevertheless in some complicated cases that involve the
calculation of the repeated integrals this recipe is insufficient.
For example, the method of \cite{T} does not allow one to
determine the expansion of an integral in a parameter if as a
result the integral itself becomes a singular function in other
parameters which, in turn, are considered as variables of the
subsequent integration. The situation becomes even more
complicated if the expanded integrand includes a product of
singular functions. Exactly this occurs in the case of pair
(multiple) production and decay of unstable particles when the
process is described basing on the expansion in the coupling
constant of Breit-Wigner factors which stand in probability (not
in amplitude) \cite {N1, N2, N-tt}. As was noted in \cite{N-tt} in
this case already in the third order of the expansion the product
of factors determined via the principal value ($V\!P$) emerges,
and giving a sense to the product is a nontrivial problem.

The purpose of the present paper is to investigate the above
mentioned situation. In narrower sense our efforts will be
directed on adding a sense to expressions of the type of an
integral in $V\!P$ sense of a singular function or of product of
singular functions if the integral itself is a singular function
of a parameter with respect to which a subsequent integration is
supposed. The problem of elimination of the ambiguities arising at
the determining of expressions of such type will be resolved by
the imposing of a simple condition certainly satisfied in the
actual applications standing beyond the present article. (The
analysis and calculations directly in the framework of the
mentioned applications see in \cite{soon}.)

The structure of this paper is as follows. In the next section we
find a solution to the problem of definition of integral in the
case when the integrand contains a single singularity regularized
by $V\!P$ prescription but the integral itself is a singular
function of integration parameters. Section \ref{VPVP} is devoted
to the determining of an integral with the product of two $V\!P$
in the integrand. In section \ref{EVP} we demonstrate the
effectiveness of the designed method on a specific nontrivial
example of calculation of integral. In section~\ref{VPn} we obtain
a generalization of the method to the case of product of three,
four and greater numbers of $V\!P$. In Conclusion the basic
outcomes of the article are specified.

\section{Integral of a single $V\!P$-pole as a distribution
over \\ integration parameters}\label{iVP}

Let us consider an integral of $V\!P$ with varying (upper) limit
of integration,
\begin{equation}\label{iVP1}
F_n(y) = \int\limits_{-\infty}^{y} \d x \; V\!P \frac{1}{x^n}\;
u(x,y)\,.
\end{equation}
Here $u(x,y)$ is a weight function usually called a test function
\cite{Sobolev,S,GS,BLOT}. It is supposed that $u(x,y)$ is distinct
from zero only in a bounded area, and is finite and differentiable
the necessary number of times.

Through well-known formula for the principal value of a pole of
degree $n$,
\begin{equation}\label{iVP2}
 V\!P \frac{1}{x^n} = \frac{(-)^{n\!-\!1}}{(n\!-\!1)!}\;
 \frac{\d ^n}{\d x^n} \ln(|x|)\,,
\end{equation}
supplied with the instruction to take an integral by the
method of ``integration by parts'' \cite{S,GS,BLOT}, one can
obtain
\begin{equation}\label{iVP3}
 F_n(y) = \frac{1}{(n\!-\!1)!}\left[
 - \!\!\!\int\limits_{-\infty}^{y} \! \d x \;
   \ln(|x|)\,u^{(n)}(x,y)
 + \ln(|y|)\,u^{(n\!-\!1)}(y,y)
 - \sum_{k=1}^{n-1} (k\!-\!1)!\>
 \frac{1}{y^k}\,u^{(n\!-\!k\!-\!1)}(y,y) \right]
\end{equation}
Here $u^{(k)}(x,y) \equiv \partial^k/\partial x^k\,u(x,y)$ and
$u^{(k)}(y,y) \equiv \partial^k/\partial x^k\,u(x,y)_{|x=y}$.

Expression (\ref{iVP3}) defines function $F_n(y)$ at any $y\not=0
$. At $y=0 $ the value of $F_n(y) $ is not determined as at $n = 1
$ expression (\ref{iVP3}) contains a logarithmic singularity, and
at $n > 1 $ a power one. Correspondingly, function $F_n(y) $ at $n
= 1 $ is integrable in a neighborhood of $y=0$, and at $n \ge 2$
is not such in the conventional sense. Nevertheless, the meaning
of function $F_n(y)$ can be extended to make it integrable in the
sense of distributions. The necessity in this operation arises if
$F_n(y)$ should be further integrated with respect to $y$.

As a recipe, a solution to the problem of extending the meaning of
a function containing a pole consists in assigning a principal
value to this pole and adding a functional concentrated at the
point of singularity of the pole. Actually the mentioned
functional must be a sum of Dirac delta function and of its
derivatives with arbitrary coefficients, and the degree of the
high-order derivative should be less by 1 than the degree of
singularity of the pole \cite{GS} (the degree of the high-order
derivative is fixed by the number of necessary subtractions in the
test function after which the integral becomes well-determined in
the conventional sense). In the case of formula (\ref {iVP3}) the
solution consists in replacing each $1/y^k$ by $V\!P\,(1/y^k) +
\sum_{l=0}^{k-1} \, C_l\,\delta^{(l)}(y)$, where $C_l$ are the
coefficients describing the parametric ambiguity.

Now we show that the ambiguities in formula (\ref{iVP3}), arising
at the extending of meaning of the poles, can be completely
removed by the imposing of a condition of independence of the
result of integration from the order of calculation of repeated
integrals. In effect, the mentioned condition means equivalence of
the result of the repeated integration to the result of the
multiple integration. (Of course, this is not the only possible
condition, but it naturally arises at solving enough wide class of
problems.)

So, let us turn to the initial formula (\ref{iVP1}) and consider
functional $\widetilde{{\cal F}}_n[u]$ specified by repeated
integration,
\begin{equation}\label{iVP5}
\widetilde{{\cal F}}_n[u] = \int_{-\infty}^{\infty} \d y
\int_{-\infty}^{y} \; \d x \; V\!P \frac{1}{x^n}\; u(x,y)\,.
\end{equation}
Our aim is to {\sl determine} this functional so that to make it
be equal to a multiple integral,\label{iii}\footnote{The
introducing of formula (\ref{iVP6}) may be considered as a
heuristic trick useful for the transition to formula (\ref{iVP7}).
Nevertheless, a precise mathematical sense can be added to
(\ref{iVP6}), as well. Really, since a singular distribution can
be considered as an improper limit of a conventional function with
respect to some parameter (by means of auxiliary regularization),
the multiple integral in (\ref{iVP6}) can be considered as a
conventional integral with the posterior transition to the limit.
Formulas (\ref{iVP6}) and (\ref{iVP7}) are equivalent from this
point of view. At the same time, functional (\ref{iVP5}) remains
uncertain as long as the transition to the limit (removal of
auxiliary regularization) before the calculation of integral $\d
y$ leads to a meaningless result. In effect we determine
functional (\ref{iVP5}) by the imposing of condition to remove
auxiliary regularization after the calculation of {\it all}
repeated integrals.}
\begin{equation}\label{iVP6}
\int\!\!\!\int\limits_{\!\!\!\!\!\!\!\!-\infty}^{\!\!\infty} \d x
\; \d y \; V\!P \frac{1}{x^n}\,\theta(y-x)\; u(x,y)\,,
\end{equation}
or to a functional determined in a similar way, but with opposite
order of the repeated integration,
\begin{equation}\label{iVP7}
{\cal F}_n[u] = \int_{-\infty}^{\infty} \d x \; V\!P
\frac{1}{x^n}\; \int_{x}^{\infty} \d y \; u(x,y)\,.
\end{equation}
It should be emphasized that in contrast to (\ref{iVP5})
functional (\ref{iVP7}) is well-determined since the integral $\d
y$ in (\ref{iVP7}) defines a function from the space of
test functions of one variable if $u(x,y)$ belongs to the space of
test functions of two variable.

By making use of the definition (\ref{iVP2}) for principal value
and carrying out obvious calculations, we obtain
\begin{equation}\label{iVP8}
{\cal F}_n[u] =-\frac{1}{(n\!-\!1)!}\int\limits_{-\infty}^{\infty}
\d x \; \ln(|x|) \left\{ \int\limits_{x}^{\infty} \d y \; \frac{\d
^n}{\d x^n} u(x,y) - \sum_{k=0}^{n-1} \frac{\d^k}{\d x^k}
\left[\frac{\partial^{\,n\!-\!k\!-\!1}}{\partial
x^{n\!-\!k\!-\!1}} \; u(x,y) \right]_{| y=x}\right\}.\nonumber
\end{equation}
In (\ref{iVP8}) all integrals are considered in a proper sense and
converge. Therefore we can change~the order of integration in the
first term. In the second term we substitute $y$ for $x$. Then we
get
\begin{equation}\label{iVP9}
{\cal F}_n[\phi] =
-\frac{1}{(n\!-\!1)!}\!\int\limits_{-\infty}^{\infty} \!\!\d y
\left\{ \int\limits_{-\infty}^{y} \!\d x \, \ln(|x|)\frac{\d
^n}{\d x^n} u(x,y) - \ln(|y|) \sum_{k=0}^{n-1} \frac{\d^k}{\d y^k}
\left[\frac{\partial^{\,n\!-\!k\!-\!1}}{\partial
x^{n\!-\!k\!-\!1}} \; u(x,y) \right]_{| x=y}\right\}.
\end{equation}
With taking into consideration (\ref{iVP2}) the later formula may
be rewritten as
\begin{eqnarray}\label{iVP10}
{\cal F}_n[\phi] \;=\!\!\!
&-&\!\!\!\frac{1}{(n\!-\!1)!}\int\limits_{-\infty}^{\infty} \!\d y
\left\{ \int\limits_{-\infty}^{y} \!\d x \; \ln(|x|)\frac{\d
^n}{\d x^n}\phi(x,y)\right.\nonumber
\\
  &-& \!\!\!\left.\ln(|y|) \frac{\partial^{\,n\!-\!1}}{\partial
x^{n\!-\!1}} \; \phi(x,y)_{| x=y} + \sum_{k=1}^{n-1}(k\!-\!1)! \;
V\!P\frac{1}{y^k}\; \frac{\partial^{\,n\!-\!k\!-\!1}}{\partial
x^{n\!-\!k\!-\!1}} \; \phi(x,y)_{| x=y} \right\}.
\end{eqnarray}
It can easily be seen that the expression in the curly brackets in
(\ref{iVP10}) coincides with that in the r.h.s. of formula
(\ref{iVP3}), considered with substitution $V\!P$-poles for the
ordinary poles and without the adding of $\delta$-functions and
its derivatives.

So, we have shown that the condition of independence of the result
of the repeated integration of $V\!P$ from the order of
calculation of the repeated integrals (the requirement of
equivalence between repeated and multiple integrations) implies
the $V\!P$ prescription for the poles arising at carrying out the
first integration in the repeated integrals. In the case when both
limits of integration are variables and when $V\!P$ is appeared
with the ``shifted'' argument, one can easily obtain the following
general formula:
\begin{eqnarray}\label{iVP11}
\lefteqn{\int\limits_{a}^{b} \d x \; V\!P \frac{1}{(x\!-\!y)^n}\;
u(x,\dots) =}
\nonumber\\
&& \frac{1}{(n\!-\!1)!}\left\{-\int\limits_{a}^{b} \!\d x \;
\ln(|x\!-\!y|)\,u^{(n)}(x,\dots) +
\ln(|b\!-\!y|)\,u^{(n\!-\!1)}(b,\dots) -
\ln(|a\!-\!y|)\,u^{(n\!-\!1)}(a,\dots)\right.
\nonumber\\
&& \qquad\qquad\>
-\left.\sum_{k=1}^{n-1}(k\!-\!1)!\left[V\!P\frac{1}{(b\!-\!y)^k}\,
u^{(n\!-\!k\!-\!1)}(b,\dots) - V\!P\frac{1}{(a\!-\!y)^k}\,
u^{(n\!-\!k\!-\!1)}(a,\dots)\right]\right\}.
\end{eqnarray}
Here $u(x,\dots) \equiv u(x,y,a,b)$ and the superscripts in
brackets mean the partial derivatives of the corresponding order
with respect to the first argument. In the case when an
integration $\d y$ is supposed as the next one, it is convenient
to represent formula (\ref{iVP11}) in the form
\begin{eqnarray}\label{iVP12}
&&\int\limits_{a}^{b} \d x \; V\!P \frac{1}{(x\!-\!y)^n}\;
u(x,\dots) =-\frac{1}{(n\!-\!1)!}\int\limits_{a}^{b} \!\d x \;
\ln(|x\!-\!y|)\,u^{(n)}(x,\dots)
\\
&&\qquad +\frac{1}{(n\!-\!1)!}
\sum_{k=0}^{n-1}\left[\frac{\d^k}{\d
y^{k}}\,\ln(|b\!-\!y|)\,u^{(n\!-\!k\!-\!1)}(b,\dots) -
\frac{\d^k}{\d
y^{k}}\ln(|a\!-\!y|)\,u^{(n\!-\!k\!-\!1)}(a,\dots)\right].
\nonumber
\end{eqnarray}
Here the derivatives with respect to $y$ are understood in the
sense of distributions, i.e. they are to be moved by the rule of
``integration by parts'' at the next integration $\d y$.

Formulas (\ref{iVP11}) and (\ref{iVP12}) admit various writings in
special cases. For example, if the dependence on $y$ in the test
function can be separated out into a factor $\phi(y)$, then there
is the following formula:
\begin{equation}\label{iVP13}
\int\limits_{a}^{b} \d x \; V\!P \frac{1}{(x\!-\!y)^n}\;
u(x,a,b)\;\phi(y) = -\frac{1}{(n\!-\!1)!}\;\phi(y)\frac{\d^n}{\d
y^{n}} \int\limits_{a}^{b} \d x \; \ln(|x\!-\!y|)\,u(x,a,b)\,.
\end{equation}
Here the derivatives are again understood in the sense of
distributions. The validity of (\ref{iVP13}) follows from the
comparison of the result of applying (\ref{iVP12}) to the
l.h.s.~of (\ref{iVP13}) with the result appearing in the
r.h.s.~after the change of integration variable $x \to x+y$ and
carrying out then the direct calculation as in conventional
integral.

The other important corollary of formula (\ref{iVP11}) appears in
the case when the test function does not depend on variable $x$
within the limits of integration $(a \dots b)$. In this case at
$n=1$ we get
\begin{equation}\label{iVP14}
\int\limits_{a}^{b} \d x \; V\!P \frac{1}{x-y} \;u(y,a,b) =
u(y,a,b)\biggl[\ln(|b-y|) - \ln(|a-y|)\biggr].
\end{equation}
At $n\ge 2$ we obtain
\begin{equation}\label{iVP15}
\int\limits_{a}^{b} \d x \; V\!P \frac{1}{(x-y)^n} \;u(y,a,b) =
-\,\frac{u(y,a,b)}{n\!-\!1} \left[V\!P\frac{1}{(b-y)^{n\!-\!1}} -
V\!P\frac{1}{(a-y)^{n\!-\!1}}\right].
\end{equation}

Let us remember once again that in formulas
(\ref{iVP11})-(\ref{iVP15}) the quantities $a$, $b$, $y$ (or, at
least, some of them) are considered as variables of a subsequent
integration but not as parameters. In the case when the mentioned
quantities are considered as parameters, the symbol $V\!P$ in the
r.h.s of the above formulas should be omitted and the formulas
themselves have a sense only at uncoincident arguments occurring
in the logarithms and/or poles.

\section{Product of two $V\!P$}\label{VPVP}

Now let us consider an integral with a more complicated structure
containing a product of two $V\!P$-poles:
\begin{equation}\label{VPVP1}
\int\limits_{a}^{b} \d x \; V\!P \frac{1}{(x\!-\!z_1)^{n_1}} \;
 V\!P \frac{1}{(x\!-\!z_2)^{n_2}} \; u(x,\dots) \,.
\end{equation}
Unfortunately, the formulas of the previous Section do not allow
one to add a sense to expression (\ref{VPVP1}) since at $z_1 =
z_2$ the integrand containing a product of singular functions is
undetermined. Nevertheless, the {\sl method} of the previous
Section, basically, can be applied in this case, as well. Really,
we can at first determine integral (\ref{VPVP1}) on the assumption
that $z_1$ and $z_2$ are the parameters of integration, not equal
one another. As a result we obtain a singular function of $z_1$
and $z_2$. Then we extend the meaning of this function in the
sense of distributions. For removal of ambiguities originating
therewith we proceed to a repeated integral of triple multiplicity
containing integration with respect to $z_1 $ and $z_2 $, and
determine this integral by the imposing of condition of
independence of the result from the order of integration. (It
should be used the property that the integral is well determined
if at first the integration is carried out with respect to $z_1$
and $z_2$, and only then with respect to $x$.)

However, the above mentioned method in the case of two poles is
found too cumbersome and hardly probable is justified, especially
if there is a greater number of $V\!P$ in the integrand. So let us
use of a trick based on an independent determination of the
product of two $V\!P$ poles. (We emphasize, once again, that the
necessity of reference to integral (\ref{VPVP1}), instead of at
once to the triple repeated integral with another order of
integration, is caused by the fact that the indicated order of
integration provides a practical solvability of a problem in some
applications.)

At first we consider the case $n_1 = n_2 = 1$, and at $x \not =
z_1 $, $x \not = z_2$, $z_1 \not = z_2$ examine the following
formula with the ordinary poles realized in the sense of
conventional functions:
\begin{equation}\label{VPVP2}
 \frac{1}{x-z_1} \ \frac{1}{x-z_2} \ = \
 \frac{1}{z_1-z_2} \; \left[\frac{1}{x-z_1} -
 \frac{1}{x-z_2}\right].
\end{equation}
The expressions in the both sides of formula (\ref{VPVP2}) may be
also considered as functionals determined on space of test
functions vanishing at the coincidence of any pair of arguments.
On such space of test functions there is still an equality
(\ref{VPVP2}) between both functionals.

Now let us state a problem about the extending of these
functionals onto the all space of test functions. The solution we
realize in two steps. At first we determine each pole by assigning
$V\!P$ and adding the $\delta$-function with arbitrary factor.
Then we determine the product of $V\!P$-poles. For solving the
latter problem we use the remarkable property of formula
(\ref{VPVP2}) which consists in the fact that its l.h.s.~may be
determined as a product of two $V\!P$ if at first the integration
is meant with respect to $z_1$ and/or $z_2$ and only then with
respect to~$x$. At the same time, the r.h.s.~of (\ref{VPVP2}) with
$V\!P$ attached to the poles is well determined if, conversely,
the integration with respect to $x$ at first is supposed. Let us
call the order of integration by regular if it coincides with the
above mentioned one, and by irregular otherwise. Then we see that
the r.h.s. of formula (\ref{VPVP2}) can be used for determining
the l.h.s.~in the case of irregular order of integration, and the
l.h.s.~can be used for determining the r.h.s. in the corresponding
case. In both cases the equating of one side of the formula to
another side should be treated as the extending of a linear
continuous functional. As was directed above, this operation is
not unambiguous and can make sense only up to a functional
concentrated at the point of singularity (uncertainty) of the
initial unextended functional. In our case this means the
necessity of adding the product of two $\delta$-functions with
arbitrary coefficient to one of the sides of the resulting
relation.\footnote{A general theory of extending of linear
continuous functionals (regularization of singular functions) is
discussed with enclosing of numerous illustrations in \cite{GS}
and \cite{BLOT}. In the context of the problem under consideration
it is appropriate to carry out an analogy with the theory of
renormalizations in the field theory: the replacing of the
ordinary poles by $V\!P$-poles and the adding to them of
$\delta$-functions can be compared with the elimination of UV
divergences accompanied by emergence of the ambiguities in
subdiagrams; the determining of the product of two $V\!P$ and the
adding of the product of $\delta$-functions corresponds to the
elimination of UV divergences and emergence of the ambiguities of
the overall type in the diagram (see Section 29 in monography
\cite{BS}).} Below we present the result obtained with the taking
into consideration of the symmetry at reading the formula from
left to right:
\begin{eqnarray}\label{VPVP3}
\lefteqn{V\!P \frac{1}{x-z_1} \ V\!P \frac{1}{x-z_2}
\;\;\mathop{=}_{def}\;\;
 V\!P\frac{1}{z_1-z_2} \; \left[V\!P \frac{1}{x-z_1} -
 V\!P\frac{1}{x-z_2} \right] }\\[0.5\baselineskip]
&& + V\!P\frac{C_1}{z_1-z_2} \;\,
 \Bigl[\delta(x-z_1)-\delta(x-z_2)\Bigr] +
 C_2 \, \delta(x-z_1)\delta(x-z_2)\,.
\nonumber
\end{eqnarray}

Here the following notes are in order. First, we have not shown
the contribution of $\delta$-function that has to be added to
$V\!P\:(z_1\!-\!z_2)^{-1} $ in the r.h.s.~since this contribution
is zero in view of nulling the expression in square brackets at
$z_1 = z_2$. Second, we have not written the contributions of
$\delta$-functions added to single poles in the l.h.s.~since by
virtue of the symmetry the structure of the relevant contributions
coincides with the structure of the second term in the r.h.s.~of
the formula. (Therefore, the mentioned contributions are absorbed
by the second term in the r.h.s.) Third, we consider the product
of $\delta$-functions in the r.h.s.~of (\ref{VPVP3}) to be
determined so that the result of its integration is independent
from the order of calculation of integrals. The latter requirement
implies that $\delta(x-z_1)\delta(x-z_2) =
\delta(z_1-z_2)\delta(x-z_2) = \delta(z_1-z_2)\delta(x-z_1)$.
Under this condition formula (\ref{VPVP3}) can be read in the
opposite direction, from right to left. In the latter case all
terms with $\delta$-functions are to be transferred to the
l.h.s.~of the relation.

Coefficient $C_1$ in (\ref{VPVP3}) can be determined through
condition of the recursive invariance of the formula. Namely, let
us demand the invariance of the formula at the removal of square
brackets in the r.h.s.~and determine the arising products of
$V\!P$ by means of the formula (\ref{VPVP3}) itself. Then we get
\begin{equation}\label{VPVP4}
C_1 = 0\,.
\end{equation}
Unfortunately, in doing so the coefficient $C_2 $ remains
uncertain, i.e. it cannot be fixed by the reasons of symmetry. So,
the determining of $C_2$ is only possible by the imposing of an
additional condition. As such condition we consider the
requirement of independence of the result of integration from the
order of calculation of the repeated integrals of the r.h.s.~of
(\ref{VPVP3}). In the case of irregular order of integration this
requirement will fix $C_2$ by equating the result of integration
to that obtained at the regular order of the calculation of
integrals. The easiest way to actualize this condition is to carry
out calculations at some special choice of the test function. The
elementary choice is the unit function in the integral with
finite~limits.

So, let us consider the following repeated integral of triple
multiplicity:
\begin{equation}\label{VPVP5}
{\cal I}_{\,x,z_1,z_2} = \int\limits_{0}^{1} \d x
\int\limits_{0}^{1} \d z_1 \int\limits_{0}^{1} \d z_2 \; V\!P
\frac{1}{x\!-\!z_1} \;
 V\!P \frac{1}{x\!-\!z_2}  \,.
\end{equation}
Its direct calculation with the use of (\ref{iVP14}) leads to
result
\begin{equation}\label{VPVP6}
{\cal I}_{\,x,z_1,z_2} = \frac{1}{3}\,\pi^2\,.
\end{equation}
On the other hand, again by direct calculating we can get
\begin{equation}\label{VPVP7}
{\cal I}_{\,z_1,z_2,x} = \int\limits_{0}^{1} \d z_1
\int\limits_{0}^{1} \d z_2 \int\limits_{0}^{1} \d x \;
V\!P\frac{1}{z_1-z_2} \, \left[V\!P \frac{1}{x-z_1} -
 V\!P\frac{1}{x-z_2} \right] = -\frac{2}{3}\,\pi^2\,.
\end{equation}
From (\ref{VPVP6}), (\ref{VPVP7}) and (\ref{VPVP3}) we conclude
\begin{equation}\label{VPVP8}
C_2 = \pi^2\,.
\end{equation}

Result (\ref{VPVP8}) can be obtained also on the basis of some
skilful manipulation with Sokhotsky formula. Really, let us
consider a product of two simple poles with the ``causal'' bypass
of singularity (see below, formula (\ref{VPVP9})). It should be
noted that this product is well determined on the space of test
functions under consideration \cite{BLOT}. If integration $\d z_1$
and $\d z_2$ at first is implied (with any test function), then
the Sokhotsky formula can be applied to each multiplier. Moreover,
in the resulting expression the brackets can be removed, as well.
Eventually we obtain
\begin{eqnarray}\label{VPVP9}
\lefteqn{\frac{1}{x - z_1 + \i 0} \; \frac{1}{x - z_2 + \i 0}
\;\doteq\;
  V\!P\frac{1}{x-z_1} \ V\!P \frac{1}{x-z_2}}\nonumber
  \\[0.5\baselineskip]
&&- \,\i\,\pi \left[V\!P \frac{1}{x - z_1} \; \delta(x - z_2) +
       V\!P \frac{1}{x - z_2} \; \delta(x - z_1)\right]
 - \pi^2 \, \delta(x-z_1)\delta(x-z_2)\,.
\end{eqnarray}
Here the point placed above the equality symbol recalls that the
equality has a sense only at the particular order of calculating
the repeated integrals.

On the other hand, provided that the integration $\d x$ at first
is carried out, the expression in the l.h.s.~of (\ref{VPVP9}) can
be transformed to the form (see the first note below
formula (\ref{VPVP3}))
\begin{equation}\label{VPVP10}
 V\!P\frac{1}{z_1-z_2} \,
 \left[ \frac{1}{x - z_1 + \i 0} -
 \frac{1}{x - z_2 + \i 0}\right].
\end{equation}
Again by applying Sokhotsky formula and removing the brackets we
get
\begin{eqnarray}\label{VPVP11}
\lefteqn{\frac{1}{x - z_1 + \i 0} \; \frac{1}{x - z_2 + \i 0}\;
=_{_{\!\!\!\!\!\mbox{\normalsize .}}}}\nonumber
  \\[0.5\baselineskip]
&&
  V\!P\frac{1}{z_1-z_2} \, \left[V\!P \frac{1}{x-z_1} -
 V\!P\frac{1}{x-z_2} \right]
 - \i \pi \: V\!P \frac{1}{z_1 - z_2}\,
 \Bigl[\delta(x - z_1) - \delta(x - z_2)\Bigr]\,.
\end{eqnarray}
The point under the equality symbol in (\ref{VPVP11}) recalls that
here another (particular) order of calculation of the repeated
integral is supposed.

Now by virtue of independence from the order of calculation of the
repeated integrals of the product of ``causal'' factors, we equate
the r.h.s.~in (\ref{VPVP9}) and (\ref{VPVP11}). Then again we
obtain (\ref{VPVP8}).

So, ultimately formula (\ref{VPVP3}) takes the form of
\begin{equation}\label{VPVP12}
V\!P \frac{1}{x-z_1} \ V\!P \frac{1}{x-z_2} =
 V\!P\frac{1}{z_1-z_2} \; \left[V\!P \frac{1}{x-z_1} -
 V\!P\frac{1}{x-z_2} \right] + \pi^2 \delta(x-z_1)\delta(x-z_2)\,.
\end{equation}
The meaning of this formula is as follows: its r.h.s.~defines the
l.h.s.~in the case of irregular order of calculation of integrals,
at first $\d x$ and only then $\d z_1$ and/or $\d z_2$. This
definition provides the equality of the result of integration of
the r.h.s.~to that obtained at the regular order of integration of
the l.h.s., at first $\d z_1$ and/or $\d z_2$ and only then $\d
x$. (It should be emphasized, once again, that the imposing of
another condition may change the value of~$C_2 $.)

A generalization of formula (\ref{VPVP12}) to the case of the
product of poles of an arbitrary degree can be easily generated
through the use of relation
\begin{equation}\label{VPVP13}
 V\!P \frac{1}{(x-z_1)^{n_1}} \; V\!P \frac{1}{(x-z_2)^{n_2}} =
 \frac{1}{(n_1\!-\!1)!\,(n_2\!-\!1)!} \>
 \frac{\d ^{n_1-1}}{\d z_1^{n_1-1}}\>
 \frac{\d ^{n_2-1}}{\d z_2^{n_2-1}}
 \left[V\!P \frac{1}{x-z_1} \; V\!P \frac{1}{x-z_2}\right].
\end{equation}
By substituting (\ref{VPVP12}) into (\ref{VPVP13}) we obtain after
simple calculation
\begin{eqnarray}\label{VPVP14}
V\!P \frac{1}{(x-z_1)^{n_1}} \; V\!P \frac{1}{(x-z_2)^{n_2}} =
 \frac{\pi^2}{(n_1\!-\!1)!\,(n_2\!-\!1)!} \;
 \delta^{(n_1-1)}(x-z_1)\delta^{(n_2-1)}(x-z_2)&&\nonumber
\\[0.3\baselineskip]
 + \sum\limits_{k=0}^{n_1-1} \mbox{$ {n_2+k-1 \choose k}$}\,(-)^k\,
  V\!P \frac{1}{(z_1-z_2)^{n_2+k}}\>
  V\!P\frac{1}{(x-z_1)^{n_1-k}}
  \qquad \qquad &&\nonumber
\\
 + \sum\limits_{k=0}^{n_2-1} \mbox{$ {n_1+k-1 \choose k}$}\,(-)^k\,
  V\!P \frac{1}{(z_2-z_1)^{n_1+k}}\>
  V\!P\frac{1}{(x-z_2)^{n_2-k}}\,.\!\!\!
  \qquad \qquad &&
\end{eqnarray}

Returning to integral (\ref{VPVP1}) introduced in the beginning of
this Section, we see that it can be determined by means of formula
(\ref{VPVP14}) and then can be calculated with the aid of the
formulas of the previous Section.

In conclusion of the present Section we note that formula
(\ref{VPVP12}) in effect is not completely new. In particular, in
\cite{Musk} a similar formula was derived, presented in the form
of a relation between repeated integrals of double multiplicity,
and in \cite{Bass} an equivalent up to notation formula was
presented (but without derivation and comments). However the
derivation in \cite{Musk}, which is based on calculation of the
conditional limits with respect to parameter of the conventional
integrals, substantially differs from our derivation which is
based on the extending of linear continuous functionals. The basic
advantage of our derivation is the universality and flexibility of
the mathematical tools in operation. This reflects, in particular,
in a possibility of automatic generalization of the results to the
case of any multiplicity of the integrals, and also in the extreme
transparence and brevity of the proposed solution.

\section{Nontrivial example of calculation of integrals}
\label{EVP}

Let us consider an example of the use of formula (\ref{VPVP12})
close to that which appears in some actual applications. Namely,
let us consider an integral over a simplex
\begin{equation}\label{EVP1}
 I(z) = \int\!\!\!\int\limits_{\!\!\!\!\!\!\!0}^{\infty}
 \d x \; \d y \;\;
 \theta(2+z-x-y)\; V\!P \frac{1}{x-1}\, V\!P \frac{1}{y-1}\,.
\end{equation}
At once we note that at $z>0 $ the point of singularity of both
poles $\{x=1,\,y=1\}$ certainly falls on the integration area. At
$z < 0$ only one of the poles can be singular. The case $z=0$ in
some sense is transitional. Simultaneously this case is
specific-singular because at $z=0$ the ambiguity of
$\theta$-function is joined to the singularity in integrand. (So,
at $z=0$ integral (\ref{EVP1}) requires of additional determining.
A possible way is noted in the footnote on page (\pageref{iii}).)

Integrals of the type of (\ref{EVP1}) arise at calculating the
probabilities of the processes of pair production and decay of
unstable particles in the approach of a modified perturbation
theory (MPT) based on expansion in the coupling constant of the
Breit-Wigner factors standing in the probability (not in
amplitude). Both $V\!P$-poles in (\ref{EVP1}) in this connection
correspond to the particular contribution emerging from the
product of two Breit-Wigner factors in the next-next-to-leading
order of the expansion of the cross-section. The points $x=1$ and
$y=1$ correspond to the positions of the mass-shells of
resonances. Quantity $z$ stands for the energy of exclusive
process counted off from the threshold of the pair production.
(See \cite{N-tt} for the establishing of correspondence, and
\cite{soon} for computation of an actual processes.)

Let us represent (\ref{VPVP12}) in the form of repeated integral
and take advantage of formula (\ref{iVP14}). Then we obtain
\begin{equation}\label{EVP2}
 I(z) = \int\limits_{0}^{2+z} \!\d x \> V\!P \frac{1}{x-1}
        \int\limits_{0}^{2+z-x}\!\!\!\d y \> V\!P \frac{1}{y-1}
        \;=\;
\int\limits_{0}^{2+z} \!\d x \> V\!P \frac{1}{x\!-\!1}\;
\ln|1\!+\!z\!-\!x|\,.
\end{equation}
Unfortunately, at $z=0$ the last integral in (\ref{EVP2}) is not
determined. Nevertheless at $z\not=0$ its calculation can be
carried out in a direct way by separating the range of integration
onto the sub-ranges. (Note that in some complicated cases this is
not always possible to make.) Omitting tiresome calculations, we
write down the result at $z>0$:
\begin{equation}\label{EVP3}
I(z) = 2\,\mbox{dilog}\!\left(1+z^{-1}\right) + \ln^2(z) -
\frac{\pi^2}{6}\,,
\end{equation}
\begin{equation}\label{EVP4}
\mbox{dilog}(z) \equiv \int_{1}^{z} \d t \; \frac{\ln (t)}{1-t}\,.
\end{equation}

The more perfect method of calculation of $I(z)$ is based on the
change of variables $x+y=\xi$, $x-y=2\eta$ (therewith the symmetry
of the going through of the integration area is achieved) and on
the usage of formula (\ref{VPVP12}):
\begin{eqnarray}\label{EVP5}
 I(z) &=& \int\limits_{0}^{2+z} \d \xi
          \int\limits_{-\xi/2}^{\xi/2} \d \eta \;\;
      V\!P \frac{1}{\eta+\xi/2-1}\;V\!P \frac{-1}{\eta-\xi/2+1}
\nonumber\\
      &=& \int\limits_{0}^{2+z} \d \xi
          \int\limits_{-\xi/2}^{\xi/2} \d \eta
      \left\{ V\!P \frac{1}{\xi-2}
      \left[
      V\!P \frac{1}{\eta+\xi/2-1}-V\!P \frac{1}{\eta-\xi/2+1}
      \right]
      - \pi^2\,\delta(\xi-2)\delta(\eta)\right\}
\nonumber\\
      &=& 2\int\limits_{0}^{2+z} \d \xi \;
          \frac{\ln|\xi-1|}{\xi-2} - \pi^2 \, \theta(z)\,.
\end{eqnarray}
The last integral in (\ref{EVP5}) can be calculated at any $z$. In
particular, at $z>-1$ we get
\begin{equation}\label{EVP6}
I(z) = -2\,\mbox{dilog}(1+z) + \frac{\pi^2}{2} - \pi^2 \,
\theta(z)\,.
\end{equation}

At $z>0$ the expressions in the r.h.s.~of (\ref{EVP3}) and
(\ref{EVP6}) are equal each other by virtue of the relation
\begin{equation}\label{EVP7}
2\,\mbox{dilog}\!\left(1+z^{-1}\right) + 2\,\mbox{dilog}(1+z) +
\ln^2(z) + \frac{\pi^2}{3} = 0\,.
\end{equation}
The validity of (\ref{EVP7}) can easily be verified by the
differentiation of the l.h.s.~with taking into consideration
(\ref{EVP4}) and by calculating separately its particular value,
for example, at $z=1$.

Thus, at $z>0$ the both above mentioned calculations are
equivalent. Nevertheless, the second method of calculation of
integral (\ref{EVP1}) allows one to solve the problem of the going
through the range of the threshold of pair production of unstable
particles, a stumbling-block from the point of view of
calculations of \cite{N-tt}. (In actual applications a singularity
arises at $z=0$, which needs in a regularization.) Besides,
outside the threshold (at $z\not=0$) the calculation by the second
method is considerably simpler, which is very important from the
point of view of the practical solvability of a problem.

\section{Product of several $V\!P$}\label{VPn}

The results of Section \ref{VPVP} may be generalized by induction
to the case of any number of $V\!P$-poles. So, the product of
three poles of degree 1 can be determined via multiplying both
sides of formula (\ref{VPVP12}) by one more pole. As a result in
the r.h.s.~we obtain the product of no more than two $V\!P$-poles
of $x$. By virtue of (\ref{VPVP12}) this product is well
determined. Furthermore, by sequential applying (\ref{VPVP12}) the
result can be reduced to the form of a sum of single poles of $x$.
Written in the completely symmetric form, the result is
\begin{equation}\label{VPn1}
 \prod_{n=1}^{3}V\!P \frac{1}{x-z_n} =
 \sum_{n=1}^{3}V\!P \frac{1}{x-z_n}
 \left[ \prod_{k\not=n} V\!P \frac{1}{z_n-z_k} -
 \frac{\pi^2}{3}\prod_{k\not=n}\delta(z_n-z_k) +
 \pi^2 \prod_{k\not=n}\delta(x-z_k)\right] .
\end{equation}

Formula for the product of four poles can be obtained by
multiplying both parts of (\ref{VPn1}) by one more pole, or by
multiplying formula (\ref{VPVP12}) by itself. Both methods after
reducing to the form of symmetrized sum of single poles of $x$
lead to result:
\begin{eqnarray}\label{VPn2}
 \prod_{n=1}^{4}V\!P \frac{1}{x-z_n}
& = &
 \sum_{n=1}^{4}V\!P \frac{1}{x-z_n}
 \left[ \prod_{k\not=n} V\!P \frac{1}{z_n-z_k} +
 \frac{\pi^2}{3}\sum_{l\not=n}
 V\!P \frac{1}{z_l-z_n}
 \prod_{{k\not=n}\atop{k\not=l}} \delta(z_n-z_k)\right]\nonumber\\
& + &\!\!
 \frac{\pi^2}{4} \sum_{P\{z_1,\dots,z_4\}}V\!P \;
 \frac{\delta(x-z_{1})\delta(x-z_{2})}{(z_{1}-z_{3})(z_{2}-z_{4})}-
 \pi^4\prod_{n=1}^{4}\delta(x-z_n)\,.
\end{eqnarray}
Here indices $k$ and $l$ run values $1,\!2,\!3,\!4$ except for the
values indicated under the symbols of sum or product. The
summation in the second term in the r.h.s.~is carried out over the
all permutations of $\{z_1,z_2,z_3,z_4\}$.

The process of multiplying by a simple $V\!P$-pole can be
continued. In doing so at each step of the induction in the
r.h.s.~we obtain the product of only two $V\!P$-poles of $x$,
which in view of (\ref{VPVP12}) is a well-determined quantity.
After the necessary number of steps we can obtain a formula that
determine the product of any number of simple $V\!P$-poles. Then
by the differentiating with respect to parameters, similarly as in
(\ref{VPVP13}), we can derive the result for the product of any
number of $V\!P$-poles of any degree. In effect this procedure is
trivial. In view of awkwardness we do not write down the ultimate
result.

\section{Conclusion}

So, if an integral in the sense of principal value of a singular
function or of product of singular functions appears by singular
function of the parameters of integration, then the integral can
be determined in the sense of distributions. The ambiguities
arising therewith are in control and can be eliminated by the
imposing of the condition of independence of the result of
integration from the order of calculation of the repeated
integrals. (The mentioned condition, of course, is not unique but
naturally arises at solving the wide class of problems.) In the
case of a single pole with $V\!P$-prescription in the integrand,
the result is described by formulas (\ref{iVP11}) and
(\ref{iVP12}). The essence of these formulas consists in assigning
$V\!P$-prescription, again, for the poles arising at the
conventional calculating of integral. The case of product of two
$V\!P$-poles in the integrand is nontrivial, but it can be reduced
to the case with a single $V\!P$-pole via the reduction formulas
(\ref{VPVP12}) and (\ref{VPVP14}). The case with many $V\!P$-poles
is easily considered by induction.

The results of the present article are extremely important for a
systematic description of the processes of pair (multiple)
production and decays of unstable particles in the high orders of
perturbation theory. Furthermore, in view of the obvious
universality they can be applied in other applications, as well,
containing the repeated integration of singular functions
determined in the sense of principal value.

The author is grateful to V.A.Petrov for valuable notes, and also
to A.I.Alekseev for the indication to Ref.\cite{Bass}, and
A.Bassetto for the indication to Ref.\cite{Musk}.

\end{document}